\documentclass[aps,prl,twocolumn,superscriptaddress,10pt]{revtex4-1}

\usepackage{amsmath}
\usepackage{amssymb}
\usepackage{wasysym}
\usepackage{graphicx}
\usepackage{hyperref}
\usepackage{color}
\usepackage{physics}
\usepackage{siunitx}

\begin{document}

\title{Rapid state-resolved single-atom imaging of alkaline-earth fermions}

\author{Thies Plassmann}
\altaffiliation{Equal contributions}
\author{Leon Schäfer}
\altaffiliation{Equal contributions}
\author{Meny Menashes}
\author{Guillaume Salomon}
\email{guillaume.salomon@uni-hamburg.de}

\affiliation{Institute for Quantum Physics, University of Hamburg, Luruper Chaussee 149, 22761 Hamburg, Germany,}
\affiliation{The Hamburg Centre for Ultrafast Imaging, University of Hamburg, Luruper Chaussee 149, 22761 Hamburg, Germany}

\date{\today}

\begin{abstract}
Local Hilbert spaces with large dimension are of key interest for quantum information with applications in quantum computing and memories, quantum simulations and metrology.  
Thanks to its weak coupling to external perturbations, the large ground-state nuclear spin manifold of fermionic alkaline-earth atoms is an exciting resource to explore for quantum information.
Simultaneous single atom and state-resolved detection however remains an outstanding challenge limiting the development of novel quantum computing and simulation schemes beyond qubits.
Here, we report on a new imaging technique enabling the simultaneous detection of up to four quantum states encoded in the nuclear spin manifold of a single fermionic strontium atom within $\sim100\, \mu s$, with state-resolved detection fidelities ranging from $93.6\%$ to $99.7\%$.
This technique is further used to track the highly coherent nuclear spin dynamics after a quench highlighting the potential of this system for quantum information.
These results offer fascinating perspectives for quantum science with multi-electron atoms ranging from qudit-based quantum computing to quantum simulations of the SU($N$) Fermi-Hubbard model.
\end{abstract}

\maketitle

Impressive progress over the last decade in the engineering, control and detection of quantum many-body systems at the level of individual constituents hold great promises for quantum computing \cite{bluvstein2024,arute2019}, quantum networks \cite{hartung2024,grinkemeyer2025}, quantum enhanced metrology \cite{finkelstein2024,cao2024} and are bringing quantum simulations to uncharted territories \cite{xu2025,qiao2025,evered2025,sompet2022,meth2025,satzinger2021}.
Advanced quantum information science platforms such as trapped ions, superconducting circuits and cold atoms trapped in optical arrays naturally allow for quantum information processing with a selected number of well-controlled quantum states -two for qubits-  or more for so-called qudits.

Trapped ions and superconducting quantum computers and simulators have recently established qudits as promising candidates for fault-tolerant computing and highlighted their potential for quantum simulations \cite{ringbauer2022,brock2025}.
Alkali-based cold atom quantum computers with single particle detection and control have so far focused on qubits and demonstrated exciting perspectives for scalability \cite{gyger2024,norcia2024,manetsch2025,chiu2025} and error correction \cite{bluvstein2026}. 
Improved qubit coherence with alkaline-earth elements \cite{barnes2022,ma2022,jenkins2022} and multilevel architectures involving both ground and excited states have been reported for quantum computing and metrology using nuclear, clock and Rydberg states \cite{ma2023,muniz2025,lis2023, pucher2024, unnikrishnan2024, ammenwerth2025, tsai2025}.

The large Hilbert space dimension $d>2$ provided by the nuclear spin of ground-state alkaline-earth fermions, with up to $d=10$ for strontium, is an attractive resource for quantum simulations and computations owing to its weak coupling to environment and $SU(N)$ invariant interactions \cite{zhang2014}.
However, quantum information beyond the two-level paradigm using more than two nuclear spin states remains an outstanding frontier in cold atom systems owing to the absence of high-fidelity state-resolved single-atom detection schemes.

Approaches for nuclear spin state detection based on narrow-line imaging, coherent momentum transfer and optical Stern-Gerlach separation have been developed for alkaline-earth bulk gases \cite{taie2010, stellmer2011,leroux2018, bataille2020}. None of these schemes have however been extended to the single atom level beyond qubits, severely limiting approaches for qudit based quantum computing \cite{kitenko2025, omanakuttan2023, omanakuttan2024,weggemans2022,buhrstein2026} and analog quantum simulations with large spin lattice fermions \cite{taie2022, pasqualetti2024,mancini2015,stuhl2015, zhou2023}. With recent developments in ultrafast imaging of cold atoms \cite{bucker2009,bergschneider2018,su2025,grun2024,muzi2025} new opportunities are emerging for single atom and state-resolved detection, paving the way for further breakthroughs in quantum information science with multi-electron atoms.

Here we report on a novel approach for rapid high-fidelity single atom and simultaneous state-resolved imaging of ground-state alkaline-earth fermions lasting $\sim 100\,\mu s$. Our technique relies on free space imaging of individual fermionic strontium atoms released from an optical tweezer in combination with a nuclear spin state dependent optical potential realizing a Stern-Gerlach separation for state-resolved detection. We first demonstrate high fidelity detection of $^{87}Sr$ in free space reaching $98\%$ for $15\,\mu s$ imaging time. Next, by flashing for $5\,\mu s$ a tightly focused laser beam detuned within the hyperfine manifold of the $^3P_1$ state we create a $|m_F|$-dependent force resulting in a discrete spatial atomic distribution after expansion at the focus of a microscope objective. State-resolved detection fidelities ranging from $93.6\%$ to $99.7\%$ for up to four $m_F$ states are reported. As an application we used this technique to track the coherent nuclear spin dynamics in the ground state after a transverse magnetic field quench, in excellent agreement with theory expectations. 
\begin{figure}[tb]
\centering
\includegraphics{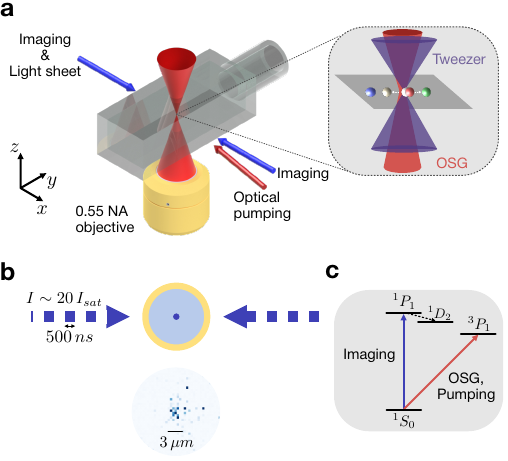}
\caption{\textbf{Experiment overview.} \textbf{a} Individual $^{87}Sr$ atoms are trapped in an optical tweezer shined through a microscope objective in a ultra-high vacuum glass cell.
Zoom in: A nuclear spin-dependent force is created by flashing 
a linearly polarized optical Stern-Gerlach beam (red, OSG) sent through the microscope objective with a slightly displaced center from the optical tweezer (violet) in the atomic plane. A subsequent free planar evolution in a light sheet dipole trap (gray plane) at the focus of the objective allows to infer the atomic spin from the atom position.
\textbf{b} Fluorescence is induced by a pair of saturating alternating beams, which are propagating along $x$ and resonant with the blue transition of $Sr$.
A few dozen photons collected via the objective are imaged on an EMCCD camera allowing to detect the presence of an atom with high fidelity (bottom experimental image).
\textbf{c} Strontium level structure. Imaging is performed on the blue transition and the optical Stern-Gerlach beam operates near the red intercombination transition. In order to initialize ground-state atoms in the nuclear spin state $m_F=+9/2$, optical pumping is performed using a $\sigma^+$ polarized laser beam on the red intercombination line, in presence of a guiding magnetic field aligned with its propagation axis \cite{supp}.}
\label{fig:fig1}
\end{figure}

Our experiments started with a few  $10^4$ $^{87}$Sr atoms trapped in a red magneto-optical trap (MOT) operating on the intercombination line at $689\, nm$ \cite{supp,mukaiyama2003}. A $813\, nm$ optical tweezer trap with waist $w_0=1.35(8)\, \mu m$ and depth $V=k_B\times 1.8\,mK$ was focused onto the atoms via a microscope objective with 0.55 numerical aperture (Fig.\,\ref{fig:fig1}). In order to enhance the atom loading efficiency in the tweezer and to confine the atom dynamics at the focus of the microscope objective for low tweezer depths, a $1040 \,nm$ elliptical dipole trap propagating along $x$ was used. An overlap of $500\,ms$ of these traps with the red MOT allowed to load a few atoms in the optical tweezer. The tweezer depth was then further ramped down in $150\,ms$ to $k_B\times 3.7\,\mu K$ resulting in one atom remaining in the trap with about 50$\%$ probability, and at a temperature $T=750(100)\, nK$.
\begin{figure}[b]
\centering
\includegraphics{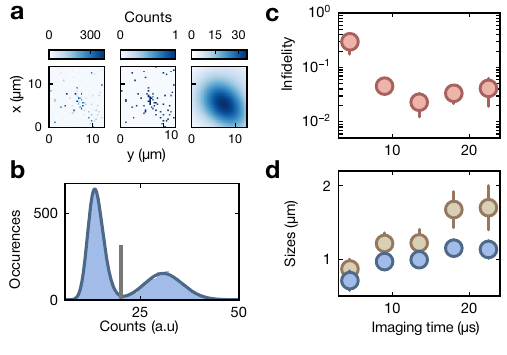}
\caption{\textbf{Rapid imaging of $^{87}Sr$ in free space.} \textbf{a} Left to right: Bias-corrected raw, binarized and low-pass filtered experimental images. \textbf{b} Histogram of the low-pass filtered images maxima, obtained for $15\,\mu s$ imaging time, displaying a characteristic double-peak structure allowing to distinguish the presence of an atom (high counts) from an empty shot (low-counts peak). A fit of the histogram (blue solid line) allows to define a threshold (vertical gray line) and to estimate the fidelity of this assignment. 
\textbf{c} Single-atom detection infidelity as a function of imaging time. \textbf{d} Spatial spread of the atomic fluorescence signal as a function of imaging time. The reported sizes are defined as the standard deviations of the fitted anisotropic gaussian along its major (yellow) and minor (blue) axis. Error bars represent one standard deviation of the mean.}
\label{fig:fig2}
\end{figure}

In a first experiment ultrafast imaging in free space for $^{87}Sr$ was developed and characterized. 
Following on a $5\,\mu s$ time of flight, two $461\,nm$ counter propagating alternating beams (Fig.\,\ref{fig:fig1}) with intensities $I\sim20\, I_{sat}$, where $I_{sat}$ is the saturation intensity on the blue transition (Fig.\,\ref{fig:fig1}c), were used to induce fluorescence for a variable time $t$. A few tens of photons collected via the microscope objective were detected on an EMCCD camera. The recorded images were binarized, low-pass filtered and the histogram of local maxima displayed a characteristic double-peak structure allowing to infer the presence of an atom (Fig.\,\ref{fig:fig2}a,b). In order to determine the optimum imaging parameters which both maximize the detection fidelity and minimize the spatial spread of the atomic signal, their respective time evolutions were studied (Fig.\,\ref{fig:fig2}c,d). Due to momentum diffusion during the scattering process the standard deviation $\sigma$ of the atomic fluorescence signal is increasing with imaging time favoring short exposures. The single-atom detection fidelity reaches a maximum at 98.2(2)$\%$ for imaging times between 13.5 and 15 $\mu s$ which we attribute to the decay channel towards the $^1D_2$ state (Fig.\,\ref{fig:fig1}), in good agreement with results reported for $^{88}Sr$ imaged in a tweezer \cite{scholl2023}.
\begin{figure*}
\centering
\includegraphics{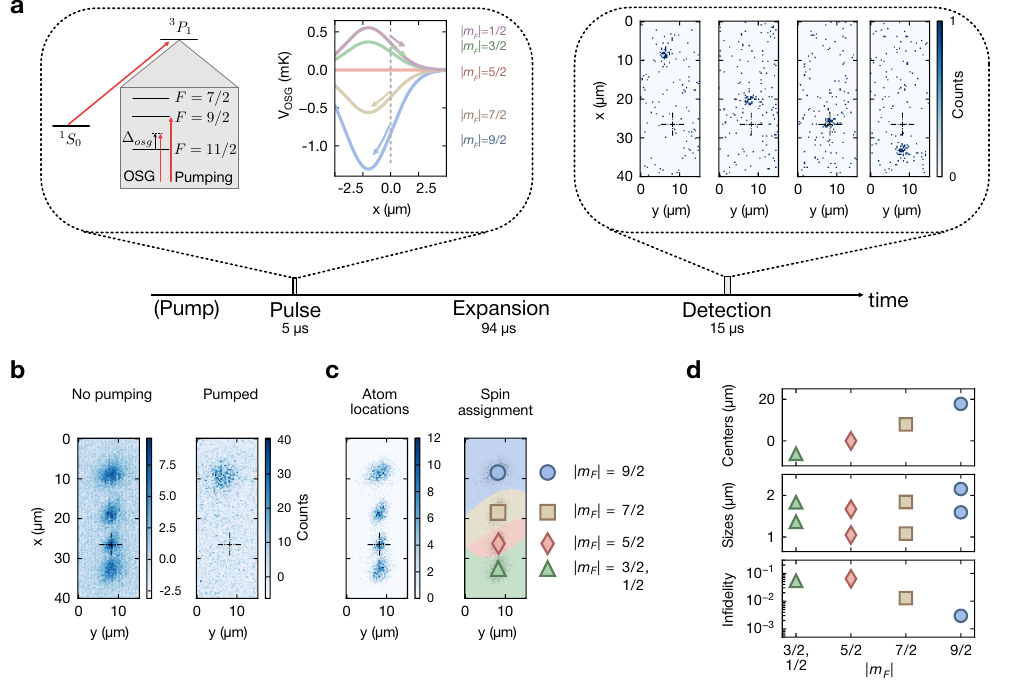}
\caption{\textbf{Spin-resolved imaging. a} An optical Stern-Gerlach beam is detuned with $\Delta_{OSG}=+790\, MHz$ from the $F=9/2\to F^\prime=11/2$ red transition creating an $|m_F|$-dependent potential gradient at the atom location (left zoom-in, vertical dashed gray line). After illuminating the atom for $5\, \mu s$ with the OSG beam, the accumulated $|m_F|$-dependent momentum change is mapped during an in-plane expansion into spatially distinct locations detected by fluorescence imaging (right zoom-in: binarized images of individual realizations). The cross indicates the tweezer position. \textbf{b} Averaged fluorescence picture with one atom (left) displaying a four-regions structure, whereas after optical pumping to $m_F=+9/2$ only one region remains (right). \textbf{c} Left: Sum of all single atom locations in 3738 experimental realizations containing an atom. The spin region assignment is determined from a four-components gaussian mixture modeling of the atom locations \cite{supp}. \textbf{d} Top: Distance of the spin regions centers from the tweezer position. Middle: Major and minor sizes of the four gaussian distributions used in the model. Bottom:  $|m_F|$ assignment infidelities.}
\label{fig:fig3}
\end{figure*}
The optimum exposure time around $15\,\mu s$ thus allowed us to reach the highest fidelity with $\mu m$-scale spatial resolution, which is of prime importance for quantum simulations and computations, and crucial for spin-resolution methods based on spatial separation. We note that fast repumping schemes for the $D$-state exist \cite{samland2024} leaving room for further fidelity improvements.

In order to simultaneously detect both the presence of an atom and the amplitude $|m_F|$ of its nuclear spin, an optical Stern-Gerlach (OSG) technique operating at the single atom level was developed. A linearly polarized laser beam blue detuned by $\Delta_{OSG}=790\,$MHz from the $F=9/2\to F^\prime=11/2$ transition on the intercombination line at $689\, nm$ was used to create a spin-dependent force immediately after release from the optical tweezer (Fig.\,\ref{fig:fig3}a). The OSG beam was sent through the microscope objective and focused down in the atom plane to a waist $w_{OSG}=4.0(2)\, \mu m$ with its center displaced from the tweezer focus by $2\,\mu m$, thereby creating a strong optical gradient at the atom position. Owing to the linear polarization of the OSG beam, aligned along $x$, the vector part of the polarizability tensor vanishes and for the chosen detuning the scalar contribution is small, resulting in a $m^2_F$-dependent force on the atoms \cite{supp,sonderhouse2020}. In this configuration the maximum number of distinguishable nuclear spin states is thus reduced from 10 to 5. After a $5\,\mu s$ pulse of the OSG beam, which is short compared to the excited state lifetime of $22\,\mu s$, the atoms evolved in the light sheet dipole trap confining their motion to the imaging plane. The averaged atomic fluorescence signal after an in-plane expansion $T_{exp}=94\, \mu s$ clearly revealed four distinct regions where the atoms can appear, with a maximum splitting between the outermost ones of $25\,\mu m$ (Fig.\,\ref{fig:fig3}b, left and Fig.\,\ref{fig:fig3}d, top). 
\begin{figure*}
\centering
\includegraphics{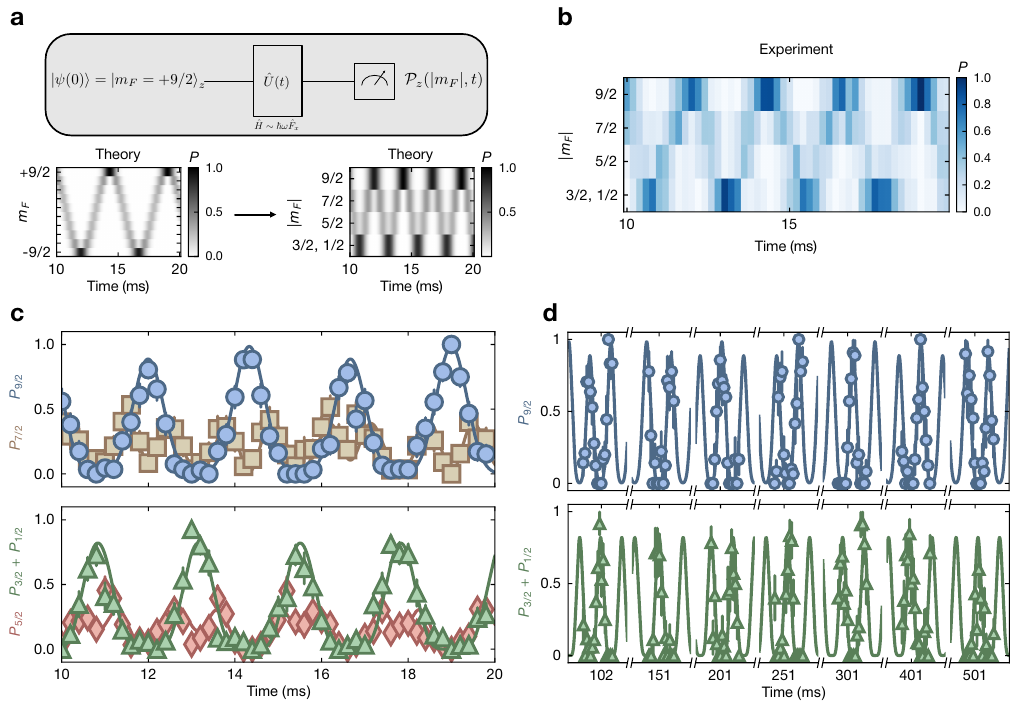}
\caption{\textbf{Coherent spin dynamics.} \textbf{a} Top: Atoms initialized in the $m_F=+9/2$ state evolve under a near-transverse magnetic field for a time $t$ before detection, which corresponds to a unitary evolution under a hamiltonian $\hat{H} \sim \hat{F}_x $ where $\hat{F}$ is a spin-$9/2$ operator. Bottom: Coherent spin dynamics theory expectations (left) induced by the field during the quench \cite{supp} and as would be measured by our detection method only sensitive to $|m_F|$ (right). \textbf{b} Experimental time evolution of the $|m_F|$ populations. \textbf{c} Time evolutions of the $|m_F|=9/2$ (blue dots), $|m_F|=7/2$ (yellow squares), $|m_F|=5/2$ (red diamonds), and $|m_F|=3/2,1/2$ (green triangles) populations and theory expectations (solid lines). \textbf{d} Long time population dynamics for  $|m_F|=9/2$ (blue dots) and $|m_F|=3/2,1/2$ (green triangles) and theory expectations (solid lines).  Error bars represent one standard deviation of the mean.}
\label{fig:fig4}
\end{figure*}
When optically pumping the trapped atoms to $m_F=+9/2$ using light resonant with the $F=9/2\to F^\prime=9/2$ intercombination transition prior to Stern-Gerlach separation \cite{supp}, only the upper region was observed (Fig.\,\ref{fig:fig3}b, right).
Given that $|m_F|=5/2$ is barely affected by the OSG beam, we could attribute $|m_F|$-states to the three regions observed from top to bottom to $|m_F|=9/2,7/2,5/2$ and the bottom one to $|m_F|=3/2,1/2$ which were not resolved for the chosen experimental parameters  \cite{supp}. Gaussian modeling of the four individual atom locations observed was used to assign region boundaries (Fig.\,\ref{fig:fig3}c) and determine spin detection fidelities (Fig.\,\ref{fig:fig3}d) of $99.70(6)\,\%$ for $|m_F|=9/2$, $98.7(2)\,\%$ for $|m_F|=7/2$, $93.6(8)\,\%$ for $|m_F|=5/2$, and $94.6(7)\,\%$ for $|m_F|=1/2,3/2$ \cite{supp}.
The different sizes of the four atom locations (Fig.\,\ref{fig:fig3}d) were attributed to a lensing effect induced by the OSG beam gaussian profile during the pulse \cite{supp}.
We occasionally detected two atoms, suggesting that our detection is compatible with initial tweezer occupations higher than one, and we leave for future work a study beyond the single atom level. 

As a benchmark for our method we tracked the coherent nuclear spin dynamics after a magnetic field quench \cite{ahmed2025}. The individual atoms were first pumped to the stretched $m_F=+9/2$ state in presence of a guiding magnetic field pointing along $x$ (Fig .\ref{fig:fig1}). 
The magnetic field was then quenched in the transverse vertical direction $z$ and held for a variable time $t$ inducing nuclear spin dynamics before detection. 
Coherent oscillations of the $m_F$ populations on a one second timescale were observed (Fig.\,\ref{fig:fig4}) in excellent agreement with theory expectations \cite{supp}.
This result both demonstrates coherent control of the nuclear spin manifold of $^{87}Sr$ at the single particle level and experimentally justifies the assignment of the $|m_F|$ regions (Fig.\,\ref{fig:fig3}b,c). The nuclear spin dynamics was found to remain coherent with no visible decay of contrast of the oscillations up to $500\,ms$ without dynamical decoupling, highlighting the potential of the nuclear spin manifold for robust quantum information storage \cite{barnes2022, ma2022, lis2023}.  

In summary we have presented here a novel rapid high-fidelity single-atom and spin resolution technique generically applicable to alkaline-earth fermions and other multi-electron atoms, which opens several avenues for quantum information science.
Combined with optical arrays with tunable spacing such as accordion optical lattices \cite{su2023} or programmable optical tweezer arrays \cite{browaeys2020}, this technique will for example allow for spin-resolved quantum gas microscopy of the SU($N$) Fermi-Hubbard model and of dipolar spin mixtures as realized with lanthanides \cite{burdick2016,trautmann2018,lecomte2025,su2023}.
Being compatible with advanced quantum computing architectures \cite{lis2023}, the reported results should further allow to leverage the large nuclear spin Hilbert space of alkaline-earth atoms for quantum information processing. In particular, it will be exciting to use it in the future to explore qudit-based quantum computing \cite{kitenko2025,omanakuttan2023,buhrstein2026}, novel mid-circuit measurements and erasure protocols \cite{scholl2023,ma2023} and to develop novel quantum error correcting codes tailored to multi-level systems.

\bigskip
\begin{acknowledgments}
\textit{Note added.} During the completion of this work, we became aware of similar results based on second-scale narrow-line imaging of $^{87}Sr$ in optical lattices \cite{icfo-private2025}.
\bigskip

\textbf{Acknowledgments:} We thank F. Scazza, L. Tarruell, S. Buob and T. Rubio Abadal for insightful discussions, and H. Moritz, S. Häfele and C. Salomon for critical reading of the manuscript. This project has received funding from the European Research Council (ERC) under the European Union's Horizon 2020 research and innovation programme (grant agreement n° 852236, "FLATBANDS"). T.P, L.S and G.S acknowledge funding support from the Cluster of Excellence "CUI: Advanced Imaging of Matter" of the German Research Foundation (DFG) - EXC 2056 - Project ID 390715994. 
\end{acknowledgments}
\bigskip
\textbf{Author Contributions:} T.P and L.S contributed equally to this work. T.P, L.S, M.M and G.S planed and conducted the study. T.P and L.S performed the data acquisition and G.S supervised the study.
All authors contributed to the analysis of the results presented here and to the writing of the manuscript.
\bigskip

\textbf{Materials and Correspondence:} Correspondence and requests for
materials should be addressed to guillaume.salomon@uni-hamburg.de

\section*{Supplementary Material}
\setcounter{figure}{0}
\renewcommand\thefigure{S\arabic{figure}}  
\subsection{Single atom preparation}
\begin{figure}
\centering
\includegraphics{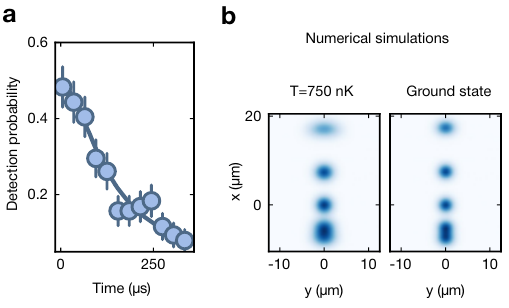}
\caption{\textbf{a} Single-atom detection probability as a function of hold time with tweezer off. Solid line: classical Monte-Carlo expectation for a temperature of $750\,nK$. \textbf{b} Numerical simulations results of the optical Stern-Gerlach detection method for a cropped thermal distribution (36 lowest energy states) in the tweezer at $T=750\,nK$ (left), and for the harmonic oscillator ground state (right), taking into account the spatial resolution of our detection due to atomic motion during imaging. As seen on the experiment only four regions are well defined and both the center distances of the $|m_F|$-regions as well as their elliptical shape are in good agreement with the experiment. From a comparison with ground-state expectations the spin-detection fidelities for the experimental settings in this manuscript were limited by the initial atomic temperature, and the absence of visible splitting between $|m_F|=3/2$ and $|m_F|=1/2$ was attributed to the atomic motion during imaging.}

\label{fig:fig1-SI}
\end{figure}

Our experimental sequence started by collecting for a few seconds $^{87}Sr$ atoms, which have a nuclear spin $I=9/2$ and hence total spin $F=9/2$ in their ground state, from a commercial cold atom beam source (AOSense) in a blue magneto-optical trap (MOT) operating on the broad $^1S_0\to\,^1P_1$ transition in an uncoated ultrahigh-vacuum (UHV) glasscell. The MOT consists of three pairs of retroreflected beams with $5\,mm$ waist and each with peak intensities of $0.3\times I^B_{sat}$ where $I^B_{sat}=40\,mW/cm^2$ is the saturation intensity on the blue transition. These beams are detuned by $\Delta=-1.2\times\Gamma_B$ from the excited state $F^\prime=11/2$, where $\Gamma_B$ is the linewidth of the 461 nm blue transition and operate in presence of a quadrupolar magnetic field with a gradient of $46\,G/cm$ along the strong $x$ axis.
During the MOT loading phase atoms accumulated in the magnetically trappable states of the $^3P_2$ manifold \cite{mukaiyama2003}. At the end of the blue MOT, repumpers operating at 707 nm and 679 nm, resonant with the $^3P_2\to6s\,^3S_1$ and $^3P_0\to6s\,^3S_1$ were turned on to pump atoms back to the ground state. Simultaneously the magnetic field gradient was ramped down to $3\,G/cm$ and three pairs of retroreflected red-detuned red MOT beams operating on the intercombination line  $^1S_0\to\, ^3P_1$ were turned on. Each beam, with $3.5\,mm$ waist and peak intensity of $3400\times I^R_{sat}$ where $I^R_{sat}=3\,\mu W/cm^2$, contains two frequencies to address both the $F=9/2\to F^\prime=11/2$ (MOT) and the $F=9/2\to F^\prime=9/2$ (stirring) transitions \cite{mukaiyama2003}. These beams were frequency modulated using a triangular shape spanning $4\,MHz$ with a $50\,\mu s$ period during a first capture phase. After $25 \,ms$ of such frequency modulation the span was linearly decreased to zero in $75\,ms$. The laser power was further reduced in 150 ms to $\sim 100\times I^R_{sat}$ and its frequency for the MOT transition ramped to a detuning $\Delta_r\simeq -650\,kHz$ while atoms started to accumulate in the $1040\,nm$ light sheet dipole trap. The light sheet dipole trap propagating along $x$ is elliptical with waists $20\,\mu m $ along $z$ and $200\,\mu m $ along $y$, and a depth of $k_\mathrm{B}\times60\,\mu K$. The $813 \,nm$ tweezer trap with 1.35(8) $\mu m $ waist and a depth $k_\mathrm{B}\times 1.8\,mK$ was shined onto the atoms from the start of the red MOT sequence. After $500\,ms$ holding time, both the red MOT beams and light sheet trap powers were ramped down to zero in 100 ms such that only atoms trapped in the tweezer remained.
In order to pump atoms to $m_F=+9/2$ a guiding magnetic field of about $2\,G$ pointing along the pumping beam propagation axis was turned on, and the tweezer power ramped down in $100\,ms$ by a factor of ten in order to decrease off-resonant scattering from the trap. Optical pumping was performed during $54\,ms$ using a frequency modulated $\sigma^+$ polarized laser beam spanning $\sim 6\,MHz$ with a $27\,\mu s$ period, and with $90\times I^R_{sat}$ peak intensity, allowing for efficient pumping in presence of excited state tensor light-shift from the tweezer trap.
In order to reduce the momentum spread of the atoms during detection, the tweezer power was reduced by another factor 50 in 150 ms, and the light sheet dipole trap was simultaneously ramped up to a depth of $k_\mathrm{B}\times6\,\mu K$ providing further longitudinal confinement at such low tweezer depth. Using a sample and hold stabilization of the tweezer power, its final depth was found to fluctuate with $2.5\%$ standard deviation.
At these final settings the tweezer trap remained filled with either a single atom or none with about $50\%$ probability. The temperature of the atoms was estimated using a release-recapture method \cite{tuchendler2008}. It consisted in switching off the tweezer for a variable time $t$ before turning it back on and measuring the probability to detect an atom using free space imaging (Fig.\,\ref{fig:fig1-SI}a). A comparison with classical Monte-Carlo predictions \cite{tuchendler2008} allowed to estimate the temperature $T=750(100)\,nK$ (Fig.\,\ref{fig:fig1-SI}a). 

\subsection{Fast imaging}
The free space imaging sequence reported in the main text was developed for strontium following on the seminal works of \cite{bergschneider2018,su2025}. Freely expanding atoms in the light sheet dipole trap were illuminated with two alternating and saturating counter-propagating beams with typical peak intensities $I=20\,I_{sat}$ in presence of a small guiding magnetic field of $80\,mG$ along $x$. The two imaging beams, linearly polarized along $y$, were turned on during $500\,ns$ in opposite phase with minimal time overlap in order to avoid fast momentum spread \cite{su2025}. 
Scattered photons from the atoms were collected through a custom microscope objective (SpecialOptics) with numerical aperture 0.55 which transmits $50\%$ of blue light. Two dichroic mirrors in the imaging path after the objective were used to split the blue wavelength from the $813\,nm$ tweezer light and the optical Stern-Gerlach (OSG) $689\, nm$ light, and two bandpass filters were used for stray light filtering in front of the sensor. The collected blue light was focused onto an EMCCD camera (Andor Ixon Ultra 897) with total magnification of 47.8 for the full imaging system.
For single photon sensitivity we operated the camera at high gain. At EM gain of 1000 we measured a gain over readout noise $g/\sigma_{read}=35$ allowing to work in the single-photon counting regime \cite{bergschneider2018}.

For further analysis the bias-corrected images were binarized using a single photon detection threshold at $6.5\times\sigma_{read}$ effectively making the readout noise of the camera negligible and minimizing the influence of the clock-induced charges (CIC) noise of the sensor, measured to be $2 \% $ per pixel for our camera settings, while keeping more than $80\%$ of the atomic photon signal \cite{bergschneider2018,su2025}.
The binarized images were further low-pass filtered taking advantage of the different spatial distributions of CIC and atomic signal using an elliptical gaussian kernel with sizes $\sigma=8,\,10.4$ pixels along its principal axis (Fig.\,\ref{fig:fig1}a). The histogram of the maxima of the binarized low-pass filtered images displayed a characteristic double peak structure corresponding to the presence or absence of an atom allowing to assign a threshold for single-atom detection (Fig.\,\ref{fig:fig1}b) and to estimate the fidelity of the imaging process. The histograms were fitted using the sum of one skewed gaussian for the peak at low counts (0-atom) and a skewed gaussian plus an offset for the 1-atom peak. The offset between the peak at low counts and the peak at high counts models the decay towards the long-lived dark state $^1D_2$ during the imaging \cite{su2025}. The optimum counts threshold for single atom assignment was obtained when the two modeled peak distributions intersects allowing to estimate a fidelity of $98.2(2)\%$ for $15\,\mu s$ imaging time. 

\subsection{Optical Stern-Gerlach}
The laser used to create the Stern-Gerlach optical potential was offset-locked to our main red MOT laser, whose frequency was stabilized onto a ULE cavity with 256 000 finesse leading to a sub-kHz linewidth. 
The OSG laser was detuned by +790 MHz from the $F=9/2\to F^\prime=11/2$ red transition and was linearly polarized along $x$. The vector polarizability contribution vanishes for linear polarization and the spin-dependent force was provided by the tensor contribution of the polarizability which scales as $m_F^2$, thereby reducing the number of different nuclear spin states which could in principle be detected from 10 to 5. For this detuning, the ground-state scalar polarizability $\alpha_s=7.2\times10^3\, a.u$ and the tensor polarizability $\alpha_t=42.5\times10^3\,a.u$. On the experiment, owing to the relatively large atomic spatial spread during imaging the distributions of $|m_F|=3/2$ and $|m_F|=1/2$ were strongly overlapping further reducing the number of different states which could be identified to four for our imaging settings (Fig.\,\ref{fig:fig1-SI}b). 
The OSG beam was projected onto the atoms via the objective to a waist of $4.0(2)\,\mu m$ and power of $2.8\,mW$. The waist of this beam was experimentally determined by using it as a red-detuned dipole trap and measuring the trapping frequency for the $^{88}Sr$ isotope. The power fluctuations of the OSG beam were kept below $2.5\%$ (standard deviation) using a sample and hold stabilization method. The relative alignment of the OSG beam and the tweezer trap was controlled by maximizing the accumulated momentum kick which is mapped in the distance from the tweezer after $94\, \mu s$ in-plane expansion.
At the start of the OSG pulse and for the free expansion, the light sheet dipole trap depth was re-increased to its maximal value in order to prevent kicking atoms out of focus.
We observed slow drifts of the relative tweezer and OSG beam alignment necessitating slight realignments on a week timescale.
In order to faithfully image the initial spin distribution it is important to align the polarization of the OSG beam and the magnetic field quantization axis prior to detection. This was achieved by optimizing the fraction of atoms detected in the $|m_F|=9/2$ region after optical pumping. 

\subsection{Analysis of spin-splitted data}
Assigning an atom to a $|m_F|$ region in an individual experimental realization with OSG detection necessitates first to define the atom location based on the fluorescence signal collected on the camera. We used the peak of the low-pass filtered images to define the atom location in a single shot \cite{bergschneider2018}.
The next step requires to model the atom locations distributions observed which reduces to a classification problem. 
Since we experimentally observed four distinct regions we fitted a gaussian mixture model \cite{scikit-learn2011} to the measured atom locations in order to estimate the probability densities of the different $|m_F|$ regions.
From the four fitted gaussian probability distributions (pdf) the optimal boundaries were defined where the pdf's are equal allowing for each realization to assign an atom to a $|m_F|$ region based on the highest pdf at this location.
The spin assignment infidelities stated in the text were estimated from the overlap of the probability density functions extracted from the gaussian mixture model distributions, and represent the sum of false positives and missed detections for each regions. The errorbars on the fidelities denote one standard error of the mean and were obtained from bootstrapping the data.
The elliptical shapes of the atom location distributions and regions centers are well reproduced from numerical simulations of the atom dynamics in the OSG beam during the pulse and its subsequent free propagation (Fig.\,\ref{fig:fig1-SI}b). The good agreement between experiment and numerics allowed to identify the temperature of the atoms as the current main limitation for the fidelities reported in this work. 

During the quench dynamics experiments, two orthogonal pairs of coils in a Helmholtz-like configuration were used to control the magnetic field. The first one defining the quantization axis prior to the quench has its strong axis aligned along $x$ and its current is kept constant during the time evolution. The second pair of coils used for the quench has its main axis aligned along $z$ and is turned on and off on a sub-$ms$ timescale for the quench.
Theory modeling of the coherent dynamics induced by the magnetic field quench is based on exact diagonalization of the dynamics of a spin $F=9/2$ in presence of the magnetic fields induced by the two pairs of coils after the state was initialized in the $m_F=+9/2$ state along the quantization axis prior to detection. We find that a quench field of $1.158\,G$ with a $1/e$ rise time of $0.3 \,ms$ and an initial guiding field of $80\,mG$ are capturing well the full experimental coherent dynamics. These fields values agree within $5\,mG$ with independent calibrations based on spectroscopy on the red intercombination line.
Stray magnetic field variations of a few $mG$ can significantly change the angle between the initial quantization basis and the detection basis provided by the OSG beam, due to the relatively weak guiding magnetic field used prior to the quench. We found that for this measurement set spanning several days an angle of 5 degree, well within experimental uncertainties, allows to reproduce well our data. 

\bibliography{bibliography}

\end{document}